\documentclass[preprint,aps,nofootinbib,showpacs,showkeys,secnumarabic,
tightenlines]{revtex4}
\usepackage{amsmath}
\usepackage{amsxtra}
\usepackage{amstext}
\usepackage{amssymb}
\usepackage{latexsym}
\usepackage{graphicx}

\begin{document}

\title{Efficiency of encounter-controlled reaction between diffusing
reactants in a finite lattice: topology and boundary effects.}

\author{Jonathan L. Bentz}
\affiliation{Department of Chemistry, Iowa State University, Ames,
Iowa 50011-3111}

\author{E. Abad}
\affiliation{Center for Nonlinear Phenomena and Complex Systems,
Universit\'{e} Libre de Bruxelles, C. P. 231, Bd. Du Triomphe,
1050 Brussels, Belgium}

\author{John J. Kozak}
\affiliation{Department of Chemistry, Iowa State University, Ames,
Iowa 50011-3111}

\author{G. Nicolis}
\affiliation{Center for Nonlinear Phenomena and Complex Systems,
Universit\'{e} Libre de Bruxelles, C. P. 231, Bd. Du Triomphe,
1050 Brussels, Belgium}

\date{\today}

\begin{abstract}
The role of dimensionality (Euclidean versus fractal), spatial
extent, boundary effects and system topology on the efficiency of
diffusion-reaction processes involving two
simultaneously-diffusing reactants is analyzed.  We present
numerically-exact values for the mean time to reaction, as gauged
by the mean walklength before reactive encounter, obtained via
application of the theory of finite Markov processes, and via
Monte Carlo simulation. As a general rule, we conclude that for
sufficiently large systems, the efficiency of diffusion-reaction
processes involving two synchronously diffusing reactants
(two-walker case) relative to processes in which one reactant of a
pair is anchored at some point in the reaction space (one walker
plus trap case) is higher, and is enhanced the lower the
dimensionality of the system. This differential efficiency becomes
larger with increasing system size and, for periodic systems, its
asymptotic value may depend on the parity of the lattice. Imposing
confining boundaries on the system enhances the differential
efficiency relative to the periodic case, while decreasing the
absolute efficiencies of both two-walker and one walker plus trap
processes. Analytic arguments are presented to provide a rationale
for the results obtained. The insights afforded by the analysis to
the design of heterogeneous catalyst systems is also discussed.
\end{abstract}

 \maketitle

\section{Introduction}
    The influence of the interplay between spatial extent and
       system  dimensionality on the reaction efficiency when reactants
       are undergoing random displacements on a finite lattice, with
       an irreversible reaction occurring on first encounter has
       attracted considerable interest over the last years \cite{JKo2000,JKo,Weiss,CNic}.
       Whereas there is a vast literature dealing with the situation
       where one of the two reaction partners is anchored at some
       point in the reaction space \cite{JKo,Weiss}, a novel feature addressed
       more recently is when \textit{both} reaction
       partners are allowed to diffuse simultaneously \cite{JKo2000,CNic}.
       This corresponds to the kinetic scheme
\begin{equation}
    \label{2wschem}
    A+A \rightarrow 2S
    \end{equation}
    where $A$ denotes a site occupied by a diffusing reactant molecule
    and $S$ a free site, as opposed to the scheme
\begin{equation}
    \label{1wschem}
    A+T \rightarrow S+T
    \end{equation}
    in which $T$ denotes a site occupied by an immobile target
    molecule (trap). In this paper, we shall
    refer to the scheme described by Eq.\ (\ref{2wschem}) as the
    ``two-walker (2W) case'',
    while the kinetics described by Eq.\ (\ref{1wschem}) will be termed
    ``one-walker plus trap (1WT) case''.

    In Refs. \cite{JKo2000,CNic}, particular
    attention was focused on processes taking place on square-planar
    lattices subject to various boundary conditions. It was found that
    significant differences in reaction
    efficiency resulted depending on whether one or both
    reactants were diffusing. The objective of the present study is to
    inquire to what extent these results are generic and, if so, how they
    depend on the geometry of the support. The latter will be characterized,
    in turn, by the size, the embedding dimension, the intrinsic
    dimensionality, and by the topological invariants. Among these, for
    two-dimensional objects (surfaces) the
    Euler characteristic $\chi=F-E+V$ where $F$ is the number of faces,
    $E$ the
    number of edges and $V$ the number of vertices is especially significant.
    In each case, the role of the boundary conditions will also be assessed.

    The mean walklength of a diffusing species before encountering
    a coreactant is a natural measure of the time scale of a
    diffusion-reaction process.  Let $\langle n \rangle_1$ be the mean
    walklength before the irreversible reaction takes place in the 1WT
    case (Eq.\ (\ref{1wschem})).  Let
    $\langle n \rangle_2$ be the corresponding quantity for the
    2W case (Eq.\ (\ref{2wschem})). The quantities
    $\langle n \rangle_{1,2}$
    are obtained by averaging over statistical realizations comprising all
    different initial configurations of both reaction partners. The smaller
    the value of $\langle n \rangle_{1,2}$, the higher the efficiency of
    the reaction.

    A point that will recur frequently in the following is that
    the reaction in the 2W
    case can occur via two different channels.  In the first scenario,
    the two reaction partners happen to occupy adjacent sites on the
    lattice, and (with a certain probability) in their next,
    mutual displacement they undergo a collision.  In the second
    channel, an intervening lattice site separates the two
    reactants and, in their subsequent motion (again, with a
    certain probability), both attempt to occupy that same (vacant)
    lattice site. As we shall see, one or both of these
    reaction channels can pertain, depending on the choice of
    boundary conditions and the parity of the total number of lattice
    sites $N$. Since both
    $\langle n \rangle_1$ and $\langle n \rangle_2$ depend on the
    characteristics of the lattice and on the boundary conditions, it will
    therefore be more appropriate to choose as measure of the relative efficiency of
    the 2W process the ratio
    $\Gamma\equiv\langle n \rangle_1/\langle n \rangle_2$.

    There are both theoretical and experimental reasons for
    studying the influence of geometry and boundary conditions on the
    efficiency of diffusion-reaction processes. First, one
    would like to isolate effects that are independent of the boundaries
    of the domain in which the diffusion-reaction processes take
    place, and for this, periodic boundary conditions are usually imposed.
    Over the last several years, however, there has been an
    avalanche of experimental work published on the study of
    diffusion-reaction processes in microheterogeneous media \cite{Barzykin}
    (micelles, clays,
    zeolites, etc.), with systems of finite extent (nanosystems)
    showing non-classical behavior (i.e., departures from ``mean
    field" behavior). Therefore, in the present study we have also
    examined in tandem with periodic boundary conditions, systems
    subject to confining boundary conditions.
    Specifically, in adopting confining boundary conditions, one
    imposes the condition that if a diffusing particle attempts to
    exit the lattice from a given boundary site, it is simply reset
    at that boundary site.

    The calculations performed in this paper are of two kinds.
    First, we use the theory of finite Markov processes to compute
    $\langle n \rangle_1$ and $\langle n \rangle_2$, for small, finite
    lattices subject either to
    periodic or confining boundary conditions.  The advantage of this
    approach is that one obtains numerically-exact solutions to the
    problem under study and, in certain cases,
    one can construct closed-form analytic solutions.  Mirroring
    this, we also perform Monte Carlo calculations, first validated
    by comparison with the Markov results
    and then used to analyze systems of large spatial extent.

    The plan of this paper is as follows. In Secs.\ \ref{D1}--\ref{d3}, we consider
    successively lattices of Euclidean dimension
    $d = 1$, $d = 2$ and $d = 3$, as well as
    the Sierpinski gasket, a lattice of fractal dimension $D = \ln3 / \ln2$.
    The influence of size effects, boundary conditions and other related
    topological features such as the connectivity
    and the Euler characteristic is discussed in each dimension. The
    main conclusions are summarized in Sec.\ \ref{conclude}.

\section{Euclidean Dimension,  $d = 1$ }
\label{D1} The 1WT problem on a $d = 1$ finite lattice subject to
periodic boundary conditions was solved by Montroll \cite{Mont}
analytically. The result he found was:
\begin{equation}
\label{1wperb}
 \langle n \rangle_1 = \frac{N(N+1)}{6}.
 \end{equation}
where $N$ is the total number of sites in the lattice.

The solution for the case of two walkers in $d=1$ for a lattice
subject to periodic boundary conditions can be obtained in closed
form by formulating the problem as a classical ruin problem and
solving the correspondence difference equations \cite{Abad}. The
result is
\begin{equation}
\label{2wperb} \langle n \rangle_2 = \left\{ \begin{array}{ll}
N(N+1)(N+2)/12(N-1) & \mbox{for $N$ even}\\
(N+1)(N+3)/12 & \mbox{for $N$ odd}\end{array}. \right.
\end{equation}
This result can also be obtained via the theory of finite Markov
processes by calculating numerically-exact values of $\langle n
\rangle_{2}$ for a series of $d=1$ lattices.  From these results
patterns can be recognized from which one can construct the above
closed-form analytic solution.

We notice a difference in the expression for $\langle n \rangle_2$
for even and odd values of $N$ also found in Refs.
\cite{JKo2000,CNic} and which will turn out to become more
pronounced in  higher dimensions. We find that this
\textit{qualitative} difference in behavior between even and odd
lattices is always manifested if periodic boundary conditions are
considered, but not if confining boundary conditions are imposed.
This behavior can be rationalized by examination of small odd and
even lattices. The key to this difference is in the reaction
mechanism, which happens either via nearest-neighbor collision
(NNC) or via same-site occupation (SSO).

\begin{figure}[h]
\begin{center}
\includegraphics[width=8.cm,height=4.cm]{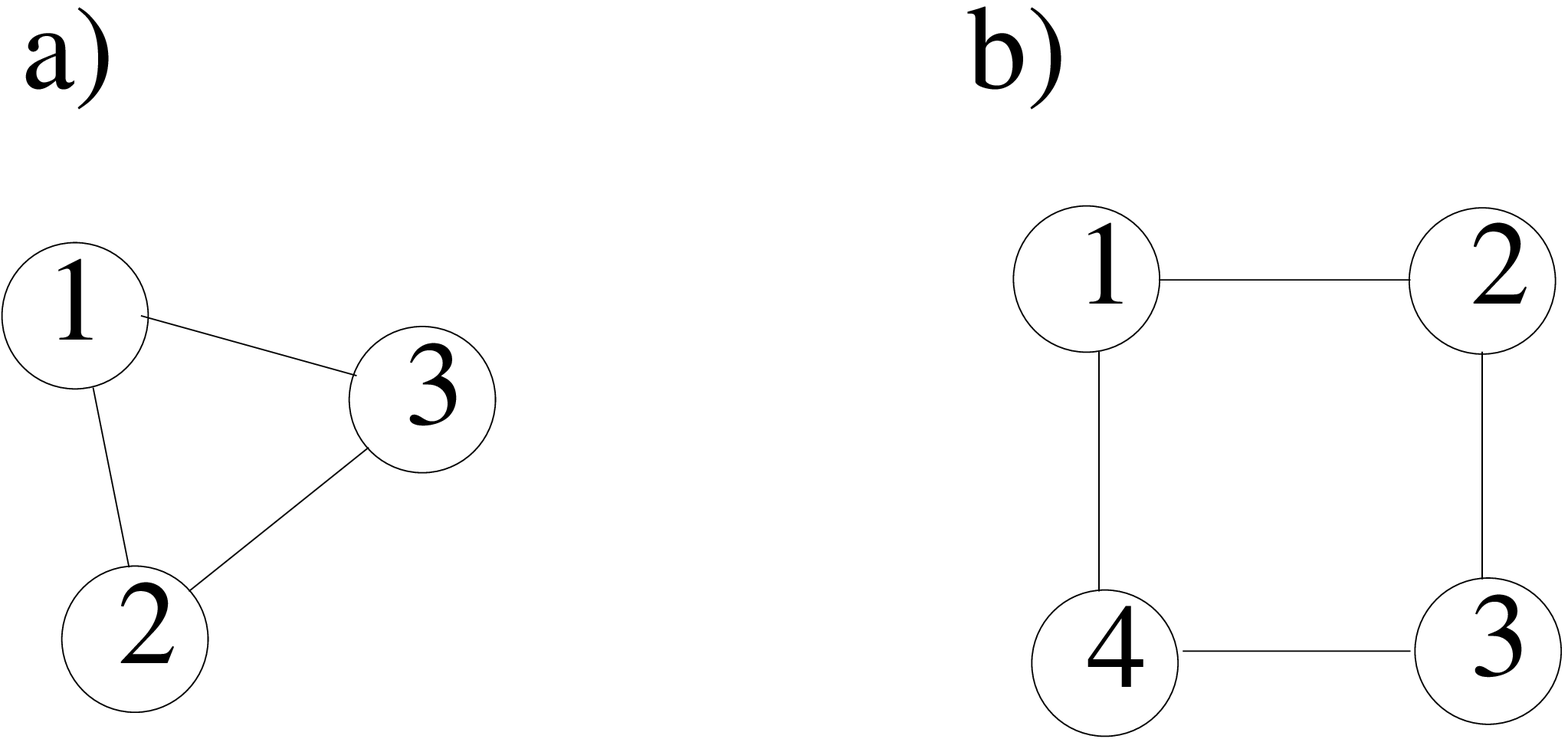}\\
\caption{}
\label{1DLattice}
\end{center}
\end{figure}

Consider the two periodic lattices of size $N=3$ and $N=4$ shown
in Figs. \ref{1DLattice}a and \ref{1DLattice}b. For $N=3$, assume
that we place a walker on site 1 and another one on site 3.
Concerned only with the concerted motions leading to reaction, one
can see that reaction takes place by SSO if both walkers jump to
site 2, or by NNC if they attempt to exchange positions. Thus,
both reaction channels are open. Now consider the periodic $N=4$
lattice and place again the walkers on sites 1 and 3. The reaction
can proceed by SSO if both walkers jump to site 2 or site 4 after
one time step. Otherwise, they will always be two sites apart,
essentially following each other in the lattice.  This means that
starting with the initial configuration of walkers on sites 1 and
3, the only allowed reaction channel is SSO. On the other hand, if
sites 1 and 2 are chosen as the initial positions of the walkers,
or more generally any adjacent pair of sites, reaction can only
occur by NNC.  This is true because any concerted motion will
either lead to NNC or the walkers will remain nearest-neighbors
after each time step. More generally, it is easily seen that for
an even lattice only one of the reaction channels is active for
any given initial configuration, whereas for odd lattices, both
can take place. This phenomenon will hereafter be referred to as
the ``even-odd effect''.

\begin{figure}[h]
\begin{center}
\includegraphics[width=8.cm,height=8.cm]{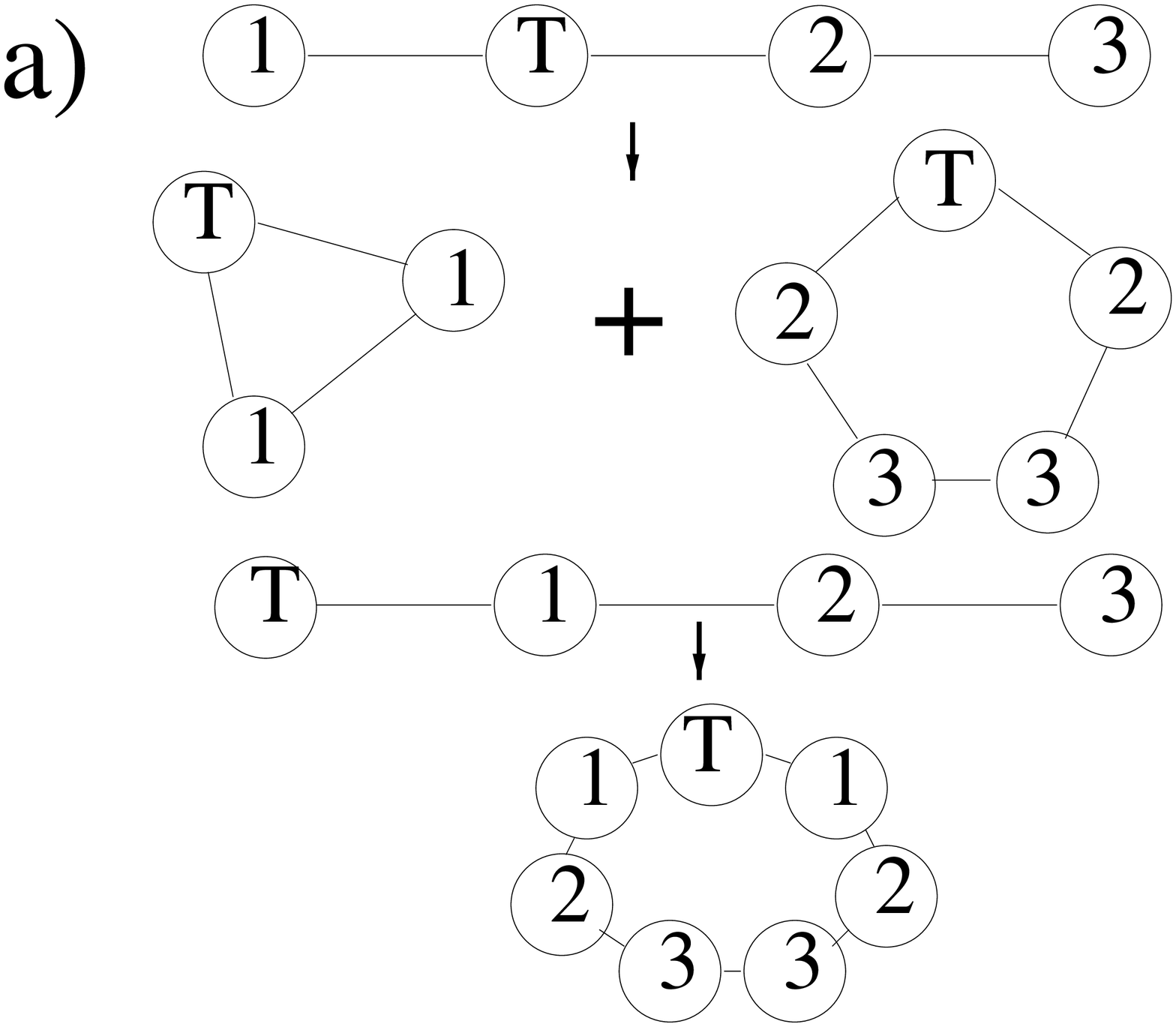}\\
\includegraphics[width=8.cm,height=1.8cm]{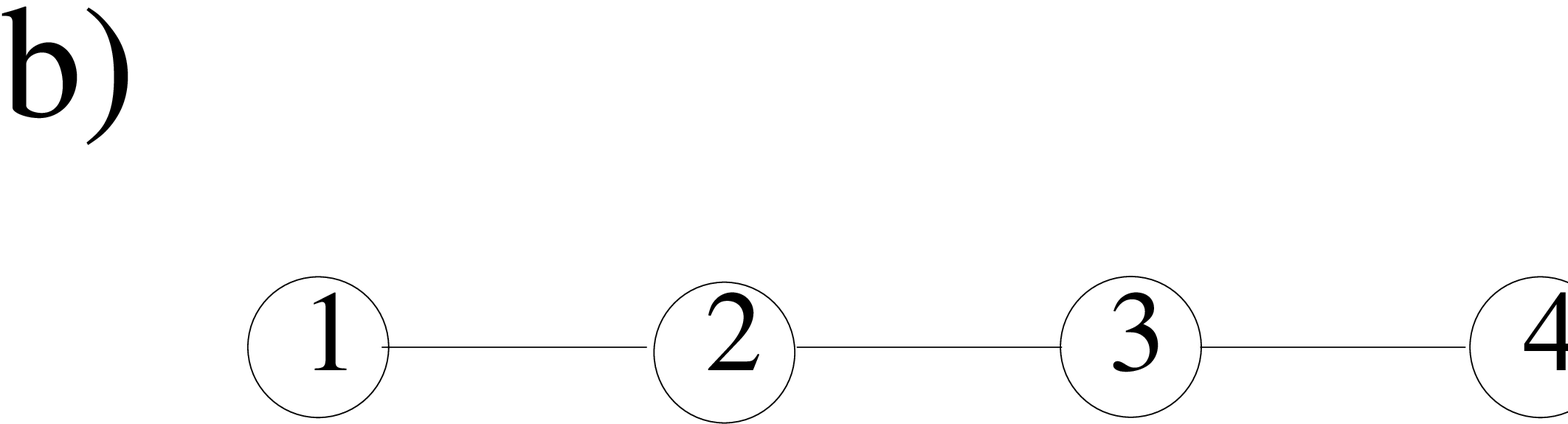}\\
\caption{}
\label{equivbc}
\end{center}
\end{figure}

We next consider the 1WT and the 2W problems in a $d=1$ lattice
subject to confining boundary conditions. From Markov theory, one
obtains for the 1WT:
\begin{equation}
 \langle n \rangle_1 = \frac{N(N+1)}{3}.
\end{equation}
According to this formula, the reaction efficiency is reduced by a
factor of 2 with respect to the periodic case described by Eq.\
(\ref{1wperb}) for all system sizes. The reason is that the
lattice with confining boundaries and a trap $T$ can be decomposed
into two disconnected periodic lattices, each of them with a trap,
since the parts of the lattice on either side of $T$ do not
communicate\footnote{For the special case where the trap is placed
at an edge site, the confining lattice reduces to a single
rather than to two periodic lattices.}.  
This equivalence is illustrated in Fig.\ \ref{equivbc}a for a
confining lattice with $N=4$. The weighted size of all equivalent
periodic lattices resulting from the different positions of the
trapping site $T$ is larger than $N$. Similar arguments are
expected to apply in higher dimensions. Thus, confinement always
reduces the efficiency of the reaction.

As for the 2W problem, no closed form expression similar to Eq.\
({\ref{2wperb}}) has yet been derived for the case of confining
boundary conditions. However,  values of $\langle n\rangle_2$ for
this case could be calculated numerically using both the Markov
method and Monte Carlo simulations.  As in the 1WT case, the
efficiency is smaller than in the periodic lattice case, but the
increase in $\langle n\rangle_2$ when confinement is imposed is
smaller than the factor of 2 found in the 1W case for all values
of the lattice size $N$ larger than two\footnote{For $N=2$, a
confining lattice is even more effective than a periodic one.}, 
 approaching a value close to 1.70 from below in the asymptotic
limit (not shown). Once more, an equivalence with the periodic
case can be established here. However, the difference with the 1WT
case is that now the size of the equivalent periodic lattices
fluctuates in time, as it depends on the instantaneous distance.
These lattice size fluctuations possibly explain why the loss of
efficiency is smaller in the 2W case when one switches from
periodic to confining boundary conditions.

We notice that no even-odd effect is to be expected in the
confining case. This is illustrated in Fig.\ \ref{equivbc}b with
two walkers initially placed on sites 1 and 3. Obviously SSO
occurs if both walkers jump to site 2. If both walkers jump to the
left, then the walker on site 3 will jump to site 2, but the
walker on site 1 will not move. Both walkers occupy now adjacent
sites and NNC
 becomes possible at the next time step. For the other symmetry-distinct
initial configuration with walkers at sites 1 and 2, one can
formulate an analogous sequence of jumps leading to SSO. Thus, for
a given initial condition, the imposition of confining boundary
conditions allows both reaction channels on an even-site lattice,
whereas with periodic boundary conditions only one of them is
realized.

\begin{figure}[h]
\begin{center}
\includegraphics[width=10.cm,height=7.cm]{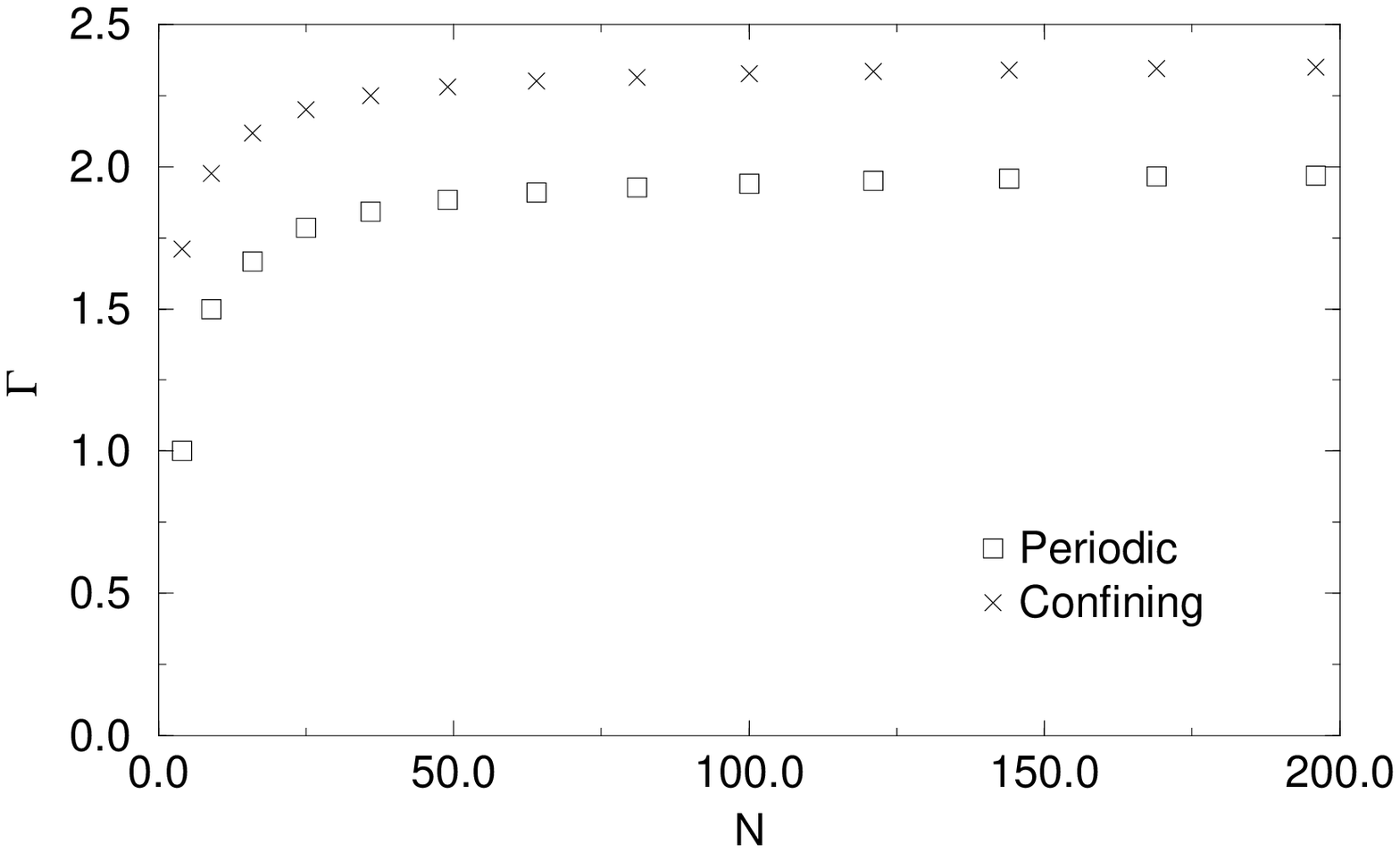}\\
\caption{}
\label{saturate}
\end{center}
\end{figure}

Figure \ref{saturate} depicts the behavior of the relative
efficiency $\Gamma$ defined in the Introduction as a function of
the lattice size $N$. It reveals an important feature, seen for
all Euclidean lattices studied regardless of dimension or class of
boundary conditions\footnote{The sole exception is the $d=3$ cubic
lattice subject to confining boundary conditions; see Section
\ref{d3a}.}. 
The curve $\Gamma$ vs. $N$ exhibits a steep increase initially
with increase in $N$ but then it ``flattens out'' and appears to
be saturating at a particular limiting value
$\Gamma_{\infty}\equiv \lim_{N\to\infty} \Gamma$. The limiting
value $\Gamma_{\infty}$ is different for different dimensions, but
the same steep increase followed by a very gradual increase to a
limiting value is found. For the periodic case in $d = 1$, the
numerical evidence suggests that the curve is approaching a value
close to 2; upon inspection of the analytic form of the solutions
given by Eqs.\ (\ref{1wperb}) and (\ref{2wperb}) one sees that the
value in the limit of large $N$ is exactly 2. This is related to
the fact that, due to the translational invariance of the lattice,
the 2W problem can be reduced to an equivalent 1WT problem, and
the effective diffusion coefficient for the latter becomes twice
as large in the large $N$ limit \cite{Abad}.

A second feature revealed by a closer examination of the analytic
form for $\Gamma$ in the periodic case is that the increase with
size is staircase-like since $\Gamma$ takes the same value for two
consecutive odd and even values of $N$ \cite{Abad}. This
consequence of the even-odd effect characteristic of the periodic
case is not seen at the scale of resolution of Fig.\
\ref{saturate} and is actually unimportant for the qualitative
behavior over sufficiently large $N$ intervals. Unlike toroidal
(periodic) lattices in higher dimensions, the even-odd effect in
one dimension does not lead to different asymptotic values
$\Gamma_{\infty}^{even}$ and $\Gamma_{\infty}^{odd}$ depending on
whether only even or only odd lattices are considered for the
computation of the respective walklengths.

\begin{figure}[h]
\begin{center}
\includegraphics[width=10.cm,height=7.cm]{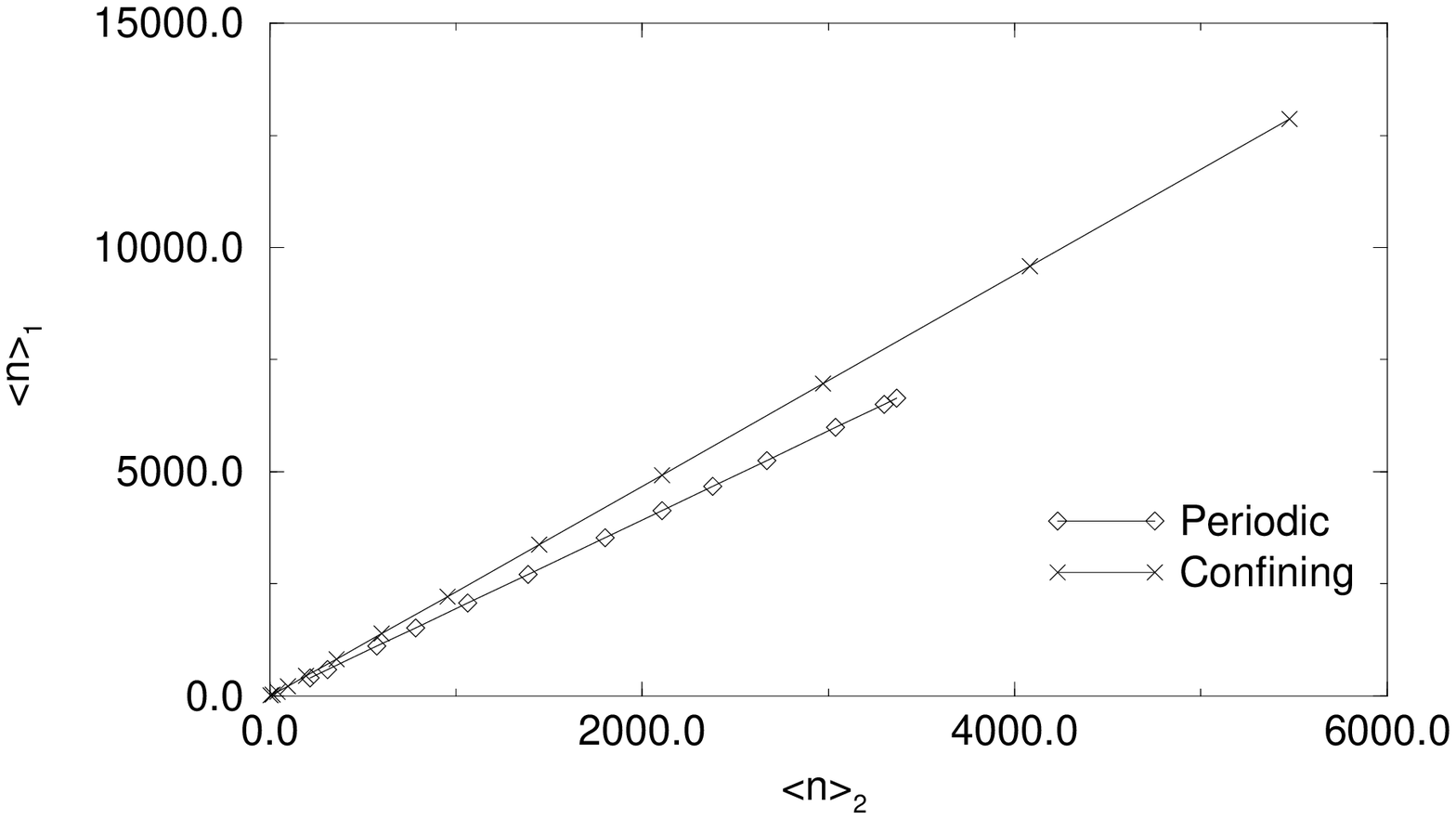}\\
\caption{}
\label{OneDResults}
\end{center}
\end{figure}

In general, plotting the $\langle n\rangle_1$ vs.\ the $\langle
n\rangle_2$ data for sufficiently large values of $N$, as
displayed in Fig.\ \ref{OneDResults}, is a convenient way of
determining the saturation value $\Gamma_{\infty}$. The slope
value from a least-squares curve fit for periodic lattices is
1.97, in good agreement with the exact analytical value, and 2.35
for confining lattices.

\section{Euclidean Dimension,  $d = 2$}\label{d2}
\begin{figure}[h]
 \begin{center}
  \includegraphics[width=10.cm,height=7.cm]{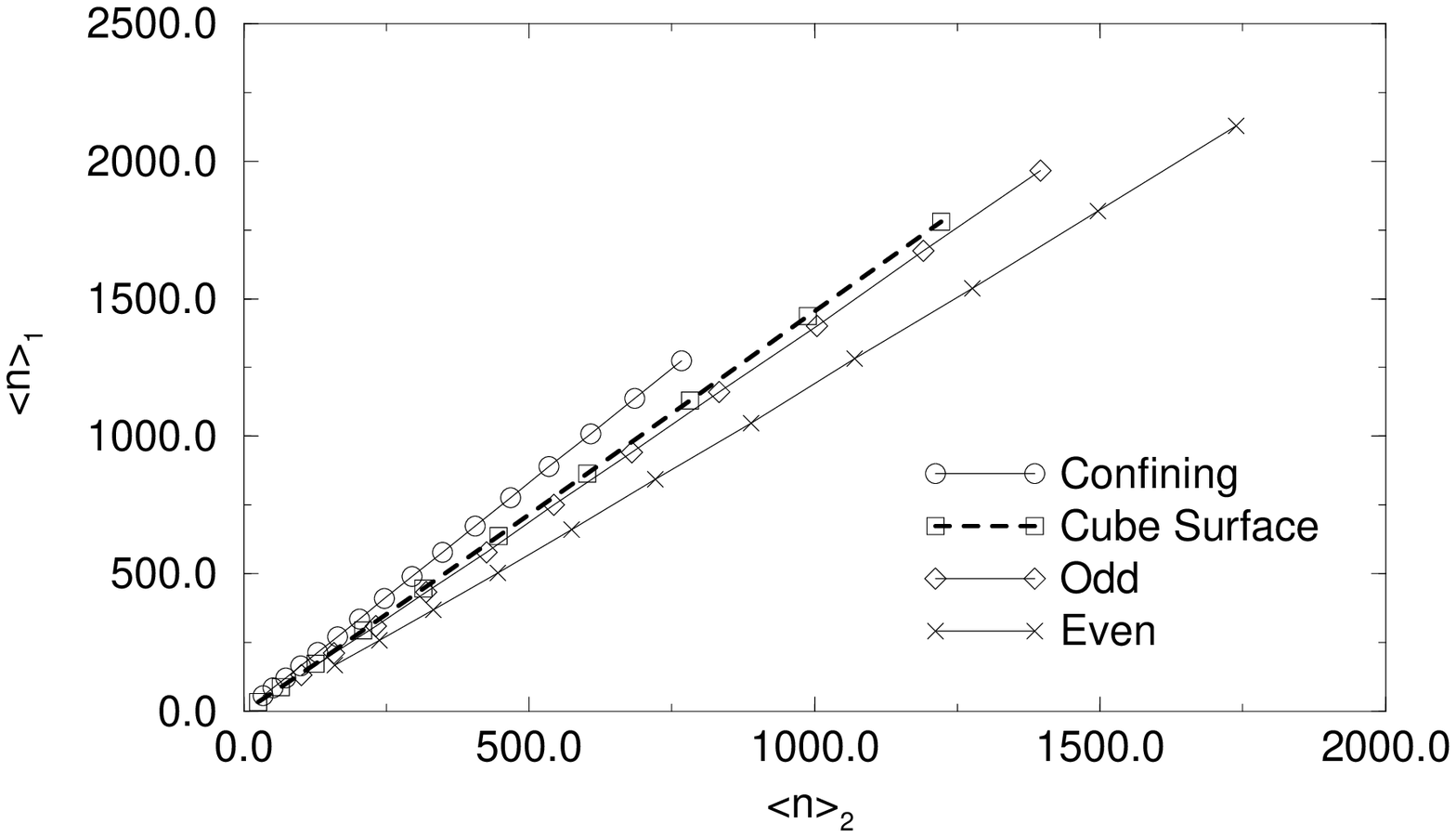}\\
  \caption{}
\label{2DPlot}
\end{center}
\end{figure}

 A square-planar lattice with periodic
boundary conditions is topologically equivalent to a torus and has
an Euler characteristic $\chi$ of 0. Results in $d = 2$ have
already been reported for the case of a square-planar lattice
subject to periodic boundary conditions \cite{JKo2000}. Figure
\ref{2DPlot} shows clearly the behavior of $\Gamma$ in the large
lattice limit, which is different for odd and even lattices.  In
Ref.\ \cite{JKo2000}, analytical evidence was provided to show
that the asymptotic limit of $\Gamma$ for odd lattices is
$\Gamma_{\infty}^{odd}=\sqrt{2}$.

As mentioned in the Introduction, surfaces are characterized
topologically not only by their Euclidean dimension but also by
their Euler characteristic.  Thus, it is also of interest to
consider non-toroidal surfaces with Euler characteristic $\chi
\neq 0$. In particular, for the sphere and the family of polyhedra
homeomorphic to it, one has $\chi=2$.  We shall consider two
examples, the first of which is the case of diffusion-reaction
processes on the surface of the Platonic solids. One of the
reasons motivating the study of such polyhedral systems is that
they mimic some features of real-world solid catalysts where the
reaction takes place on particular crystallographic faces of the
solid. In earlier work \cite{Polit1}, the vertices of the Platonic
solids were used as the allowable particle positions. The movement
of each particle was permitted along any adjoining edge. In the
present study, we report the results of the calculation using the
faces of the Platonic solids; the particle moves from face to
adjacent face, crossing only one edge at each time step. The
relationship between these two calculations is reciprocal. In
formal language, two polyhedra are dual if the number of faces of
one of them equals the number of vertices of the other, and vice
versa \cite{Henle}. Recall that the tetrahedron has 4 faces and 4
vertices and both faces and vertices have a valency of 3. The
octahedron has 8 faces and 6 vertices, and the hexahedron (cube)
has 6 faces and 8 vertices. The valency of the octahedron faces is
3 as is the valency of the cube vertices. Likewise, the valency of
the octahedron vertices is 4 as is the valency of the cube faces.
Performing the calculation on the faces of the octahedron is
equivalent to the calculation on the vertices of the cube, and the
same reciprocal relation holds true when examining the icosahedron
and the dodecahedron. Table \ref{platonic} shows results for
$\langle n \rangle_1$ and $\langle n \rangle_2$ on the Platonic
solids. As can be seen, with the exception of the octahedron,
$\Gamma$ increases with $N$, in line with the results for the
square planar lattice.

It is instructive to view the results of the Platonic solids in
tandem with their planar analogues. The tetrahedron, octahedron
and hexahedron can be placed in correspondence with periodic
planar lattices with the same valency and same value of $N$;  the
dodecahedron and icosahedron have no periodic, planar analogues.
In correspondence with the tetrahedron and octahedron we
constructed hexagonal lattices for comparison, while the
hexahedron was compared with a $2 \times 3$ square-planar lattice.
The tetrahedron and hexahedron have smaller values than their
planar analogues for the walklengths $\langle n \rangle_{1}$ and
$\langle n \rangle_{2}$ but larger values of $\Gamma$.
Surprisingly all the walklength values for the octahedron and the
8-site hexagonal lattice are identical. Upon further examination
of this degeneracy, one finds that the values are identical
because the connectivity of the octahedron is identical, site by
site, with the connectivity of the 8-site periodic hexagonal
lattice.  These calculations show, surprisingly, that the
connectivity of the lattice seems to play a relatively more
important role than the Euler characteristic in determining the
value of the mean walklength.

We also examined the influence of imposing confining boundary
conditions on the planar lattice analogues of the Platonic solids.
Owing to the fact that the connectivity of sites on the boundary
of these (finite) lattices changes, the values of $\langle n
\rangle_1$ and $\langle n \rangle_2$ change. The values
 of $\langle n \rangle_2$ for all three planar analogues are only
slightly larger than for the case where periodic boundary
conditions are imposed. On the other hand, the values of $\langle
n \rangle_1$ are considerably larger than the values of $\langle n
\rangle_1$ calculated for periodic boundary conditions; for the
$N=4$ hexagonal lattice, the $N=8$ hexagonal lattice and the $N=2
\times 3$ square-planar lattice, the value of $\langle n
\rangle_1$ increases by a factor of 3, 4 and 2, respectively.
Thus, in small systems, changes in the boundary conditions have a
significant effect on the calculation of $\langle n \rangle_1$,
but less so on the calculation of $\langle n \rangle_2$. Overall,
the influence of boundary conditions on both the 1WT and 2W cases
for these planar analogues of the Platonic solids is more drastic
than is the case for a (small) $d=1$ lattice. This is reflected in
the value of $\Gamma$, which increases by a larger factor here
than in the $d=1$ case when confining boundary conditions are
imposed.

In order to emulate large-size systems, we have studied
diffusion-reaction processes on the surface of a cube, which is
also of Euclidean dimension $d = 2$ and Euler characteristic
$\chi$ = 2. The particle is confined to the surface, and each face
of the cube is divided into a $N \times N$ square planar grid so
that there are $N \times N$ accessible sites on each face. The
total number of sites is $N\times N \times 6$. Note that for $N=1$
the previous case of an hexahedron is recovered. The valency of
each site is exactly four, keeping in mind that some movements of
the walker will take it to a different face.

Figure \ref{2DPlot} summarizes the $d = 2$ results in the large
lattice limit. It is seen that, on the surface of a cube, the
differential efficiency $\Gamma$, given by the slope of the curve
$\langle n \rangle_1$ vs.\ $\langle n \rangle_2$, is larger than
on a periodic but smaller than on a confining square planar
lattice. As expected, the efficiency of 1WT processes is once
again decreased in a more pronounced way when confinement is
imposed. The even-odd effect is not seen for processes on the
surface of a cube or for planar lattices subject to confining
boundary conditions. On the cube, an argument analogous to the one
put forth for $d = 1$ applies, except that the walkers on the cube
surface have two additional degrees of freedom. However, this has
no influence on the fact that, irrespective of the starting
configuration, both NNC and SSO may still take place because of
the the walkers' ability to traverse edges and thereby migrate to
different faces of the cube.

\section{Higher Dimensions and Fractal Structures}\label{d3}
\subsection{Euclidean Dimension,  $d = 3$  } \label{d3a}

\begin{figure}[h]
  \begin{center}
  \includegraphics[width=10.cm,height=7.cm]{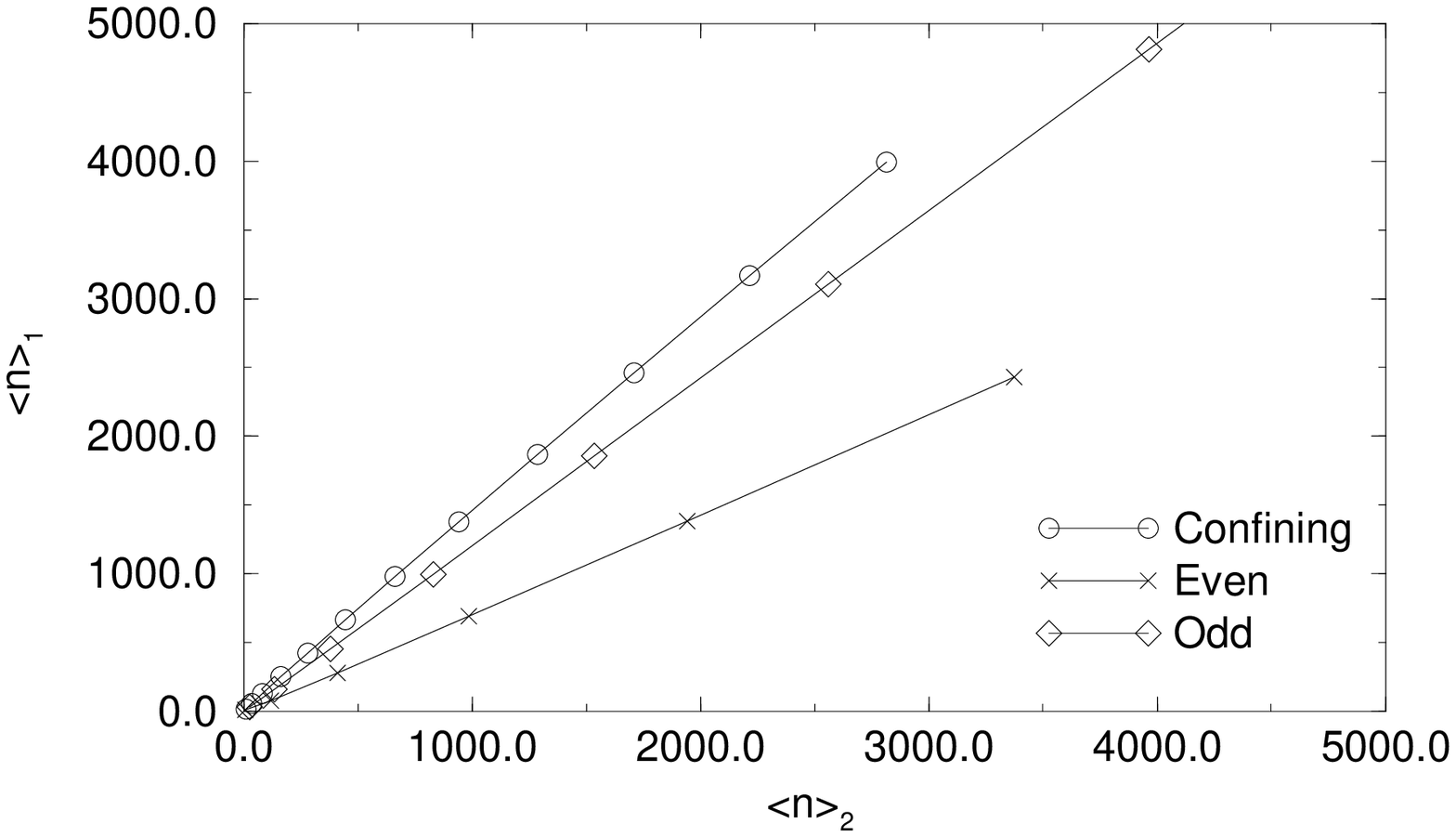}\\
  \caption{}
\label{3DOpen}
  \end{center}
\end{figure}

Numerically-exact results have been reported previously
\cite{CWalsh2,CWalsh1} for the 1WT case on a $d = 3$ cubic lattice
subject to periodic boundary conditions. We have extended this
study to calculate $\langle n \rangle_1$ and $\langle n \rangle_2$
for both periodic and confining boundary conditions by means of
Monte Carlo simulations. In $d = 3$, the even-odd effect is even
more pronounced than in the square-planar lattice case, as may be
seen in Fig.\ \ref{3DOpen}. From the least-square fits we infer
$\Gamma_{\infty}^{even}=0.72$ and $\Gamma_{\infty}^{odd}=1.22$.
This suggests that the even-odd effect is enhanced with increasing
dimensionality of the hypercubic lattice. Note also that
$\Gamma_{\infty}<1$ for even lattices, the only case where the
differential efficiency goes below unity except for the
octahedron.

\begin{figure}[h]
\begin{center}
  \includegraphics[width=10cm,height=7cm]{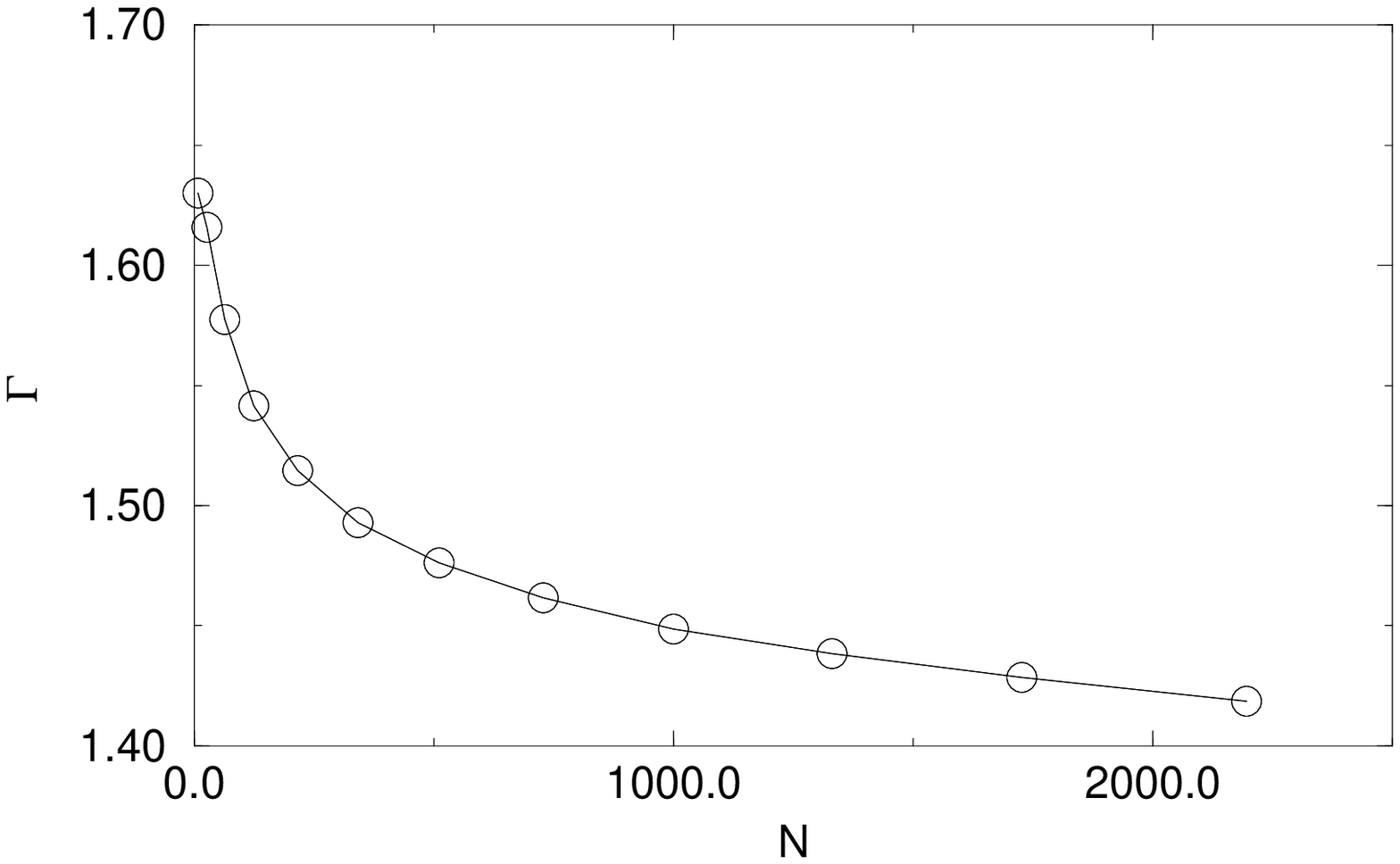}\\
  \caption{}
\label{3DCon}
\end{center}
\end{figure}

For confining boundary conditions, another significant result for
$d = 3$ is shown in Fig.\ \ref{3DCon}. As is evident, $\Gamma$
approaches its limiting value \textit{from above}, and
$\Gamma_{\infty}$ is found to be about 1.42 in this case. In all
previous cases, namely the $d = 1$ and $d = 2$ cases with both
types of boundary conditions, the graphs had the general features
of Fig.\ \ref{saturate}, approaching $\Gamma_{\infty}$ from below.

\subsection{Fractal Dimension, $D =\ln3 / \ln2$  }
From the results presented for the cases $d=1$, the $d=2$
square-planar lattice, and $d=3$ cubic lattice subject to
confining boundary conditions, one can infer that, with increasing
dimensionality d of the lattice, $\Gamma_{\infty}$ decreases
monotonically taking the values 2.35, 1.66 and 1.42, respectively,
for $d$=1, 2 and 3. In order to assess further the role of
dimensionality and of confinement, one can next consider the case
of a Sierpinski gasket, whose ramified self-similar structure is
described by an intrinsic non-integer dimension $D= \ln3/\ln2
\simeq 1.585$. The gasket can be constructed hierarchically as a
limit of successive generation gaskets. Each generation gasket is
characterized by an index $i$. The $i=0$ generation is an
equilateral triangle whose 3 vertices play the role of lattice
sites, i.e.\ allowable reactant positions. The $i=1$ generation
can be constructed by appending to each of the bottom vertices of
the primary triangle an additional equilateral triangle so that
the upper vertices of the appended triangles are identical with
the bottom vertices of the primary one and the appended triangles
have a common bottom vertex. The $i=2$ generation is constructed
by performing the same procedure on the resulting structure, i.e.
appending twice the same structure at its bottom, and so on. Thus,
at a given generation step, the number of identical substructures
increases by three while the linear size doubles, yielding the
above value of $D$ for the Sierpinski gasket obtained when
$i\to\infty$. For each generation, the lattice sites are
identified with the three apex vertices located at the outmost
corners of the corresponding gasket and with the common vertices
of any pair of adjacent triangles in the gasket. The number of
sites in the $i$th generation gasket is $N=N(i)=(3/2)\,(3^i+1)$.

The gasket shares some common features with a square-planar
lattice. First, the embedding dimension is the same, viz. $d=2$.
Secondly, except for the vertex sites  on the gasket, the valency
of all other sites is $v=4$; increasing the size of the gasket
decreases the fraction of lattice sites not having a valency of 4.
All sites on a square-planar lattice subject to periodic boundary
conditions have a valency $v=4$. On imposing confining boundary
conditions on the square-planar lattice, all interior sites will
have valency $v=4$, vertex sites will have valency $v=2$ and
boundary sites will have valency $v=3$; the percentage of vertex
and boundary sites on a square-planar lattices also decreases with
increase in lattice size. Thus, both for the gasket and for a
finite square-planar lattice subject to confining boundary
conditions, the relative importance of interior sites on the
statistics increases with increasing lattice size. Finally, one
can easily convince oneself that in the 2W case both the SSO and
NNC channels can take place for any given configuration of the
walkers on the gasket, so no even-odd effect is expected here
either.

In view of the above resemblances, it is interesting to inquire to
what extent the results obtained for the relative efficiency
$\Gamma$ on the gasket are similar to those for the square lattice
with confining boundary conditions. To this end, $\langle n
\rangle_1$ and $\langle n \rangle_2$ and their ratio were
calculated using the theory of finite Markov processes for $N$ =
6, 15, and 42 and via Monte Carlo simulations for all generations
up to $N$= 3282. For the 1WT case, a closed-form analytic
expression for the walklength is already available if the trap is
maintained at one of the apex vertices of the $n$th generation
gasket, namely \cite{Bala}
\begin{equation}
\label{n1apex} \langle n \rangle_{1,a}= \frac{3^i\,
5^{i+1}+4\,(5^i)-3^i}{3^{i+1}+1}.
\end{equation}

In performing Monte Carlo simulations, the amount of computing
time can increase dramatically within increase in gasket size $N$
thus limiting the number of statistical realizations $n_{real}$.
In the 1WT case, each realization comprises all sets of possible
configurations for the walker and the trap, while in the 2W case,
each realization comprises all sets of possible configurations for
the two indistinguishable walkers. The analytic expression of
$\langle n \rangle_{1,a}$ given in Eq.\ (\ref{n1apex}) was used to
test the accuracy of MC results for gasket sizes $N>123$, where
the Markovian approach becomes cumbersome due to the size of the
relevant transition matrices. It was found that, as $N$ is
increased, the number of realizations required to reach a given
accuracy
 in $\langle n \rangle_{1,a}$
with respect to the exact result given by (\ref{n1apex})
 decreases significantly. For all values of $n_{real}$ given in
Table \ref{sierpinski}, the relative error turned out to be
$<0.4\%$ (data not shown). Therefore, the results for $\langle
n\rangle_{1}$ and $\langle n\rangle_{2}$ for large $N$ displayed
in Table \ref{sierpinski} seem reliable enough to describe
correctly the qualitative behavior of the efficiencies.

Although the convergence to a hypothetical saturation value is
very slow in $N$, one notices from Table \ref{sierpinski} that
$\Gamma$ appears to exceed the value 2 with increasing $N$, in
contrast to the result $\Gamma_{\infty}=1.66$ for a confining
square lattice. Assuming that $\Gamma$ keeps on increasing
monotonically with the gasket size $N$, a question of interest is
whether $\Gamma$ approaches a saturation value less than the value
$2.35$ found for a $d=1$ confining lattice, or whether the
asymptotic value remains between the values $1.66$ and $2.35$, as
one might expect in view of the monotonic decrease of
$\Gamma_{\infty}$ with decreasing dimensionality referred earlier.

\section{Conclusions}\label{conclude}
In this work, results, both analytic and numerical, on the role of
boundary effects and of geometric factors such as size,
dimensionality and topological invariants on the efficiency of
encounter-controlled reactions have been obtained. Values for the
mean walklengths $\langle n\rangle_1$ and $\langle n\rangle_2$ for
the 1WT and the 2W case as well as the relative efficiency
$\Gamma$ have been computed both for large lattices and small size
systems.

The relevance of considering in detail systems of restricted
spatial extent is increasing, as it is nowadays realized that
heterogenous catalytic processes of great importance take place on
single crystallographic faces of solid catalysts where they can
involve only a few tens of particles. Despite the fact that the
numerical values obtained in this work for small systems are not
universal, some universal trends of a different kind have
nevertheless been observed, e.g.\ the dependence (increasing,
monotonic, etc.) of $\Gamma$ on valency, connectivity and boundary
conditions. In general, one has $\Gamma>1$, implying that the
reactive efficiency of two moving reactants is greater than a
diffusing plus an immobile one.

Turning now to large-size systems, our results show that $\Gamma$
increases with increasing lattice size until it reaches a
well-defined limiting value $\Gamma_{\infty}$. For a given lattice
geometry, this value decreases with increasing dimensionality.
With the exception of the even cubic lattice, $\Gamma_{\infty}>1$.

For Euclidean lattices of square-planar or cubic symmetry and
subject to periodic boundary conditions the value of
$\Gamma_{\infty}^{odd}$ is exactly 2 and $\sqrt{2}$ in
one\footnote{In the diffusive limit, it can be shown that the
diffusion coefficient for the relative motion of both reactants is
twice as large in the 2W case \cite{Abad}. Discrepancies from this
value for small systems are due to the discreteness of the
lattice. Possibly similar arguments hold in higher dimensions.} 
 and two dimensions respectively, and about 1.22 in three
dimensions. One is tempted to advance that this last number is
actually within the precision afforded by the simulations just
$\sqrt{(3/2)}$. Now, 1, $\sqrt{2}$ and $\sqrt{3}$ are the natural
metrics of the lattices here considered in, respectively, 1, 2 and
3 dimensions. The results for $\Gamma_{\infty}^{odd}$ could then
be particular cases of a universal expression: the ratio of the
maximum distance that two simultaneously moving reactants can
traverse in one time unit before they react or find themselves in
nearest neighbor positions, over the analogous quantity for the
one reactant plus trap case. Further work is necessary to assert
the validity of this conjecture and to understand why it manifests
itself only for odd lattices.

For a given system size, the 1WT and the 2W reaction efficiency
becomes less efficient when confinement is introduced, but the
decrease in efficiency is smaller for the 2W case, leading to an
increased value of $\Gamma$ with respect to the periodic case.
This boundary effect appears to become less significant with
increasing system size, although it is not completely absent in
the thermodynamic limit.

The questions raised in this work and the results obtained
constitute potentially useful elements in the important problem of
optimal design of the microreactors nowadays involved in chemical
kinetics under nanoscale conditions. For instance, as seen in
Sec.\ III, a small catalytic surface in the form of a sphere or of
a hexahedral surface homeomorphic to it ($\chi=2$) would enhance
the reaction efficiency as compared to a surface homeomorphic to a
torus ($\chi=0$). These observations highlight the need to
incorporate in the design such aspects as the geometry of the
microreactor, which can enhance an increasingly effective mixing
of the reactants and hence an increased efficiency of the reaction
itself. Finally, the role of the kinetics (linear vs.\ nonlinear)
in modulating or enhancing the importance of such factors is
certainly a problem worth addressing and this study is underway.

\begin{acknowledgements} It is a pleasure to thank Dr. M. Plapp for
valuable suggestions concerning the hierarchical construction of
the Sierpinski gasket. This work is supported by a NATO
Cooperative linkage grant PST.CLG.977780 and by the European Space
Agency under contract number 90043.
\end{acknowledgements}

\newpage

\subsection*{Figure Captions}
\noindent Figure 1:  Periodic $d=1$ lattices with a) $N=3$ and b)
$N=4$.

\noindent \\Figure 2:  a) Correspondence between confining and
periodic boundary conditions for the two symmetry-distinct trap
configurations in a confining 4-site lattice. Sites with same
numbers are symmetry-equivalent. b) Four site confining lattice.

\noindent \\Figure 3:  $\Gamma$ vs.\ $N$ in $d = 1$.  The curve
for the periodic case is generated from the exact analytical
results and the data for the confining case is from Monte Carlo
simulations.

\noindent \\Figure 4:  $\langle n \rangle_1$ vs.\ $\langle n
\rangle_2$ in $d = 1$. The slopes of the best-fit lines are:
Periodic = 1.97 and Confining = 2.35.  $N_{\mathrm{min}}=49$ for
the periodic case ($N_{\mathrm{min}}$ is the smallest data point
used in the calculation of the best fit curve.)  Both curves have
R values of greater than 0.999.

\noindent \\Figure 5:  $\langle n \rangle_1$ vs.\ $\langle n
\rangle_2$ in $d = 2$. The slopes of the best-fit lines are:
Confining = 1.66, Even = 1.25, Odd = 1.42, Cube Surface = 1.47.
All data are calculated from simulations.  The values of
$N_\mathrm{min}$ are: Confining = 25, Even = 100, Odd = 81, Cube
Surface = 24.  All R values are greater than 0.999.

\noindent \\Figure 6:  $\langle n \rangle_1$ vs.\ $\langle n
\rangle_2$ in $d = 3$. The slopes of the best-fit lines are: Odd =
1.22, Even = 0.72, Confining = 1.42.  All data are calculated from
simulations.  $N_\mathrm{min}=8$ for all the curves with all R
values greater than 0.999.

\noindent \\Figure 7:  $\Gamma$ vs.\ $N$ in $d = 3$ subject to
confining boundary conditions; all data are calculated from
simulations.

\newpage

\newpage
\begin{table}[htb]
  \centering
\begin{tabular}{c|ccccc}
\hline \hline
  Object & $N$ & valency & $\langle n \rangle_1$ & $\langle n \rangle_2$ &
$\Gamma$ \\
\hline
  Tetrahedron & 4 & 3 & 3 & 3 & 1 \\
  Hexagonal Lattice & 4 & 3 & 3.6667 & 4.8214 & 0.7606 \\
\hline
  Octahedron & 8 & 3 & 8.2857 & 9 & 0.9206 \\
  Hexagonal Lattice & 8 & 3 & 8.2857 & 9 & 0.9206 \\
 \hline
  Icosahedron & 20 & 3 & 28.8421 & 21.0955 & 1.3672 \\
\hline
  Hexahedron & 6 & 4 & 5.2 & 5.0182 & 1.0362 \\
  Square Lattice & 6 & 4 & 5.7714 & 5.8915 & 0.9796 \\
  \hline
  Dodecahedron & 12 & 5 & 12.7273 & 11.1676 & 1.1397 \\
  \hline\hline
\end{tabular}
\caption{Analytic results for Platonic solids and planar analogues
with periodic boundary conditions}\label{platonic}
\end{table}

\begin{table}[htb]
  \centering
\begin{tabular}{cc|ccc|cccc}
  \hline\hline
&  &\multicolumn{3}{|c}{Markov Results} & \multicolumn{4}
  {|c}{Monte Carlo Results} \\
 $i$ & $N$ & $\langle n \rangle_1$  & $\langle n \rangle_2$ &
  $\Gamma$ &
  $\langle n \rangle_1$ & $\langle n \rangle_2$ &
  $\Gamma$ & $n_{real}$ \\
  \hline
 1 & 6 & 6.5 & 4.3860 & 1.4820 & 6.500 &  4.386  & 1.482 & $10^7$\\
 2 & 15 & 25.6857 & 16.0957 & 1.5958  & 25.686 & 16.096 & 1.596 &$10^7$\\
 3 & 42 & 118.0582 & 68.9220 & 1.7129  & 118.06 & 68.94 & 1.713 &$10^6$\\
 4 & 123 &-  & - & - & 578.18 & 315.90 & 1.830  & $10^5$\\
 5 & 366 & - & -  & - & 2886.1 & 1500.9  & 1.923 & $10^3$\\
 6 & 1095 & -& - & - & 14465.3 & 7282.9 & 1.986 & 10\\
 7 & 3282 &- & - & - & 72512.6 & 35724.1 & 2.029 & 1 \\
  \hline\hline
\end{tabular}
\caption{ Walklength results for successive Sierpinski generation
gaskets.} \label{sierpinski}
\end{table}


\begin{thebibliography}{widest-label}
\bibitem{JKo2000} J. J. Kozak, C. Nicolis, G. Nicolis,
{J. Chem. Phys.} 113 (2000) 8168.

\bibitem{JKo} J. J. Kozak, {Adv. Chem. Phys.} 115
(2000) 245.

\bibitem{Weiss} G. H. Weiss, {Aspects and Applications of the Random
Walk,} North-Holland, New York/Amsterdam, 1994.

\bibitem{CNic} C. Nicolis, J. J. Kozak, G. Nicolis,
{J. Chem. Phys.} 115 (2001) 663.

\bibitem{Barzykin} A. V. Barzykin, K. Seki, M. Tachiya, {Adv.
Coll. Inter. Sci.} 89-90 (2001) 47.

\bibitem{Mont} E. Montroll, {J. Math. Phys.}  10
(1969) 753.

\bibitem{Abad} E. Abad, G. Nicolis, J. L. Bentz, J. J. Kozak, Physica A
(submitted).




\bibitem{Polit1} P. A. Politowicz, J. J. Kozak, {Proc.
Natl. Acad. Sci. USA.} 84 (1987) 8175.

\bibitem{Henle} Michael Henle, {A Combinatorial Introduction to
Topology,} W.H. Freeman and Company, San Francisco, 1979.

\bibitem{CWalsh2} C. A. Walsh, J. J. Kozak, {Phys. Rev. B.}
 26 (1982) 4166.

\bibitem{CWalsh1} C. A. Walsh, J. J. Kozak, {Phys. Rev. Lett.}
47 (1981) 1500.

\bibitem{Bala} J. J. Kozak, V. Balakrishnan, {Phys. Rev. E.}
65 (2002) 021105.



\end{thebibliography}
\end{document}